\documentclass[useAMS,usenatbib,usegraphicx,onecolumn]{mn2e}
\begin{document}
\title[Hadronic Models for {\rm Fermi}-LAT GRBs]{Hadronic Models for LAT Prompt
  Emission Observed in {\it Fermi} Gamma-Ray Bursts} 
\author[P. Crumley and P. Kumar]{P.~Crumley,$^1$\thanks{E-mail: crumleyp@physics.utexas.edu, pk@surya.as.utexas.edu}
 P.~Kumar,$^2$\footnotemark[1]\\
$^1$Physics Department, University of Texas at Austin, Austin, TX 78712\\
$^2$Astronomy Department, University of Texas at Austin, Austin, TX 78712
}

\maketitle
\begin{abstract}
This paper examines the possibility that hadronic processes produce
the $\ga$100 MeV photons in the prompt phase of gamma-ray bursts
(GRBs) observed by the \textit{Fermi}-LAT. We calculate analytically
the radiation from protons and from secondary electron-positron pairs
produced by high energy protons interacting with gamma-rays inside of
the GRB jet. We consider both photo-pion and Bethe-Heitler pair
production processes to create secondary electrons and positrons that
then radiate via inverse Compton and synchrotron processes. We also
consider synchrotron radiation from the protons themselves. We
calculate the necessary energy in protons to produce typical {\it
  Fermi}-LAT fluxes of a few $\mu$Jy at 100 MeV. For both of the
photo-pion and Bethe-Heitler processes, we find that the required
energy in protons is larger than the observed gamma-ray energy by a
factor of a thousand or more. For proton synchrotron, the protons have
a minimum Lorentz factor \(\sim 2\times10^6\). This is much larger
than expected if the protons are accelerated by relativistic
collisionless shocks in GRBs.  We also provide estimates
  of neutrino fluxes expected from photo-hadronic processes. Although
  the flux from a single burst is below IceCube detection limits, it
  may be possible to rule out photo-hadronic models by adding up the
  contribution of several bursts. Therefore, photo-hadronic processes
seem an unlikely candidate for producing the {\it Fermi}-LAT radiation
during the prompt phase of GRBs.
\end{abstract}

\section{Introduction}
The {\it Fermi} satellite has detected 29 gamma-ray bursts (GRBs) with
photons of energies \(\ga\)100 MeV with the Large Area Telescope
(LAT) \citep{2009ApJ...697.1071A}. The photons detected during the prompt phase of GRBs by the
{\it Fermi}-LAT are emitted with a delay of a few seconds for long
GRBs, and are observed for a longer duration of time compared to the
lower-energy photons (\(\sim\)1 MeV) observed by the {\it Fermi}
Gamma-ray Burst Monitor (GBM) \citep{2009Sci...323.1688A}. Hadronic
models could explain the delayed emission detected by the {\it
  Fermi}-LAT because of the additional time needed to energize the
protons \citep{Razzaque10,Asano2012}.  Since protons do not suffer from
radiative losses like electrons, it is possible that GRBs accelerate
protons much more efficiently, causing the hadronic emission to
dominate at high energies, \citep{Asano09,Murase12}.  The relative
efficiency of accelerating protons or electrons in GRBs depends on the
mechanism that accelerates the particles. Two likely mechanisms for
particle acceleration in GRBs are shock acceleration
\citep[\textit{e.g.}][]{Bell78,Blandford78,Blandford87} or magnetic
dissipation
\citep[\textit{e.g.}][]{1992Natur.357..472U,2002A&A...391.1141D,2003astro.ph.12347L,2006NJPh....8..119L}. In
this work, instead of dealing with the detailed specifics of particle
acceleration for realistic GRB jet models, we assume that high energy
protons exist as a power-law distribution and calculate the energy
required in protons to reproduce the high-energy photon flux observed
in the LAT band during the prompt phase.

There is a growing body of evidence that the observed emission in the
LAT band after the prompt phase may be due to synchrotron radiation
from the early afterglow emission, where electrons from the medium
surrounding the GRB are accelerated by the external forward shock
\citep{Kumar09,Gao09,Corsi10,Kumar10}. The most convincing evidence of
external forward shock producing the $\ga 100$ MeV photons is that the
late time optical and X-ray afterglow data can accurately predict the
earlier flux in the LAT band \citep{Kumar09}. Under the external
forward shock origin for the LAT emission, the delay in the LAT is due
to the delayed onset of the external forward shock emission. The
external-forward shock emission begins when the jet emitted by the
central engine deposits half of its kinetic energy to the surrounding
medium. Therefore, there remains the possibility that while the later
time ($T\ga T_{90}$\footnote{$T_{90}$ is the time at which the GRB has
  radiated 90\% of its observed photon energy}) LAT-band flux is from
the early afterglow, the earliest high energy photons observed during
the prompt phase are created by a different process. The external
forward shock is unable to explain the claimed variability present in
the LAT prompt emission \citep{Maxham11}. Furthermore, although the
most GRB spectra can be fit with a smoothly-joined broken power law
extending several decades in energy \citep[\textit{i.e.} the Band
  function,][]{1993ApJ...413..281B}, a few bursts exhibit deviations
from the simple Band function. Notably, GRBs 090902B, 090926A, and
090510 show evidence of an additional high-energy spectral component
\citep{Zhang:1292911,Fermi090902B,Fermi090926A}. This spectral feature
is usually transient and disappears before the prompt phase is
over. For example, in GRB 090510, which has a $T_{90}$ of 1.5 s, it
becomes statistically impossible to distinguish the additional
power-law component $\sim0.8$ s after the trigger
\citep{Asano09}. Similar spectral evolution was seen in GRB 090926A,
where an additional power law component appears after $\sim11$ s, and
$T_{90}$ is 13.6 s \citep{Fermi090926A}. This suggests that the
high-energy prompt emission may have a different origin than the
extended high-energy emission for some GRBs.

This paper is similar to \citet{Asano09} and \citet{Asano10} in that
it addresses interactions between high-energy protons and the
low-energy (GBM) photons as a possible mechanism to produced the high
energy photons observed in {\it Fermi}-LAT detected GRBs, but it
differs in a few major ways. First, this paper investigates hadronic
models of radiation---{\it i.e.} photo-pion production, Bethe-Heitler
pair production, and proton synchrotron not only to explain the extra
spectral components of the few GRB that exhibit them, but also to
investigate the possibility that hadronic models are responsible for
the majority of the GRB prompt emission observed in the {\it
  Fermi}-LAT. Second, instead of using Monte Carlo methods or detailed
numerical codes, we use simplifying assumptions to calculate
everything analytically or semi-analytically whenever
feasible. Finally, the most important point of this paper is to
provide analytical estimates of the energy required in protons to
explain the LAT flux with hadronic emission. These estimates allow us
to see how the energy requirement depends on various parameters.

The LAT and GBM detected photons are likely to have separate origins
during the prompt phase because of the measured delay between
them. Since we are interested in explaining the observed LAT flux of
GRBs, we will assume that the $<100$ MeV portion of the GRB emission
is of unknown origin and perfectly described by a Band function fit
with an observed peak energy of $\nu_p \sim 1$ MeV (the exact value,
is of course, different for different bursts, and \(\nu_p\) also evolves
with time during a single burst). The Band function has a low energy index
$\alpha$ and a high energy index $\beta$.  We only consider
interactions between the photons from the Band function and
high-energy protons in the GRB jet. That is, we do not consider
second-order processes, such as proton colliding with a pion-produced
positron, et cetera. This choice greatly decreases the computational
difficulty but produces results that only hold in regions where the
contributions from these processes are likely small, {\it i.e.} above
the photosphere.  In two zone models, the lower energy
  emission is produced at a significantly smaller radius than the
  radius where high energy emission is produced. As \citet{zoupiran11,
    2012MNRAS.421..525H} have noted, a two zone model will reduce the
  minimum Lorentz factor by a factor of $\sim2-5$ compared to the
  value obtained using a one zone model such as
  \citet{lithwick01,2009Sci...323.1688A,2009A&A...498...89G}. A two
  zone model is unlikely to change our results by a significant
  factor. We provide the dependencies of the efficiency on Lorentz
  factor to help determine if a lower Lorentz factor could make the
  energy requirement more attractive.

The paper is organized as follows: In section \ref{pion} we calculate
the positrons produced through the photo-pion process. In section
\ref{pion_est} we provide an analytical estimate for the required
luminosity in protons to match a flux of 1 $\mu$Jy at 100 MeV with the
photo-pion process and provide the corresponding
  neutrino flux in \ref{pion_nu}. The required proton luminosity
calculation is then repeated in section \ref{pion_better} however it
is done numerically and more accurately. In addition, the maximum
efficiency of the photo-pion process is calculated for given GRB
parameters. In section \ref{Bethe_Heitler}, we calculate the electrons
produced through Bethe-Heitler pair production and compare this to the
photo-pion calculation in \S\ref{pion_better}. Section \ref{pro_synch}
contains a calculation of the maximum efficiency for the proton
synchrotron production of typical fluxes for LAT detected GRBs,
as well as an expected neutrino flux from the proton
  synchrotron model. Finally, in section \ref{conclusions} we
summarize and discuss our results.  For calculations of the luminosity
distance we take cosmological parameters of \(H_0=71\ {\rm
  km/s/Mpc},\ \Omega_m=0.27,\ \Omega_\Lambda=0.73\).

\section{Photo-pion Production}\label{pion}

The photo-pion process refers to the production of pions ($\pi^0$ and
$\pi^\pm$) in collisions of photons with protons. The decay of these
pions produces high energy electrons and positrons that can then
produce high energy photons via synchrotron radiation. The photo-pion
process is likely to be important in situations where electrons are
unable to be accelerated efficiently to very high Lorentz factors, but
protons are. It also offers a way to beat the well known limit on the
maximum synchrotron photon energy of $\sim 50\frac{\Gamma}{1+z}\ {\rm
  MeV}$ for shock-accelerated electrons, where \(\Gamma\) is the
Lorentz factor of the source and $z$ is the redshift
\citep{1996ApJ...457..253D}.

The delta resonance, $p^++\gamma\rightarrow\Delta^{+}$, has the
largest cross section and has the lowest energy threshold of the
photo-proton resonances and is therefore the most important photo-pion
interaction to consider. The delta resonance has two main decay
channels, $\Delta^+\rightarrow\pi^++n$ and $\Delta^+\rightarrow
\pi^0+p^+$. The neutral pions quickly decay further as
$\pi^0\rightarrow \gamma+\gamma$.
%\rightarrow e^++e^-$.  
The threshold photon energy for photo-pion production is
  approximately 200 MeV in the proton rest frame. The photon energy at
  the peak of the GRB spectrum, in jet comoving frame, is
  $\nu_p(1+z)/\Gamma$, where the peak frequency in the observer frame is $\nu_p$
  and the GRB jet is moving with a Lorentz factor $\Gamma$.  For a
  proton to undergo pion production with photons of energy $\nu_p$,
  the proton Lorentz factor must satisfy
\begin{equation}\label{eq:min_lf_p}
\gamma_p\ga 2\times10^4\Gamma_2\nu_{p,6}^{-1}(1+z)^{-1},
\end{equation}
where $\nu_{p,6}$ is the observed peak frequency in units of $10^6$ eV; the
standard notation $X_n\equiv \frac{X}{10^{n}}$ is used. At the
threshold, the $\pi^0$ is produced more or less at rest with respect to
the proton. This means that the $\pi^0$ will decay into two photons
with energies \(\sim 100\ \Gamma^2_2\nu_{p,6}^{-1}(1+z)^{-2}\) TeV in
the observer rest frame. This energy is well outside the {\it
  Fermi}-LAT band. The high-energy photons created through $\pi^0$
decay will interact with the lower energy photons to produce $e^\pm$
pairs. If the optical depth of the \(\gamma+\gamma\) pair production
for 1 TeV photons is much greater than one, they will readily produce
$e^\pm$ pairs with Lorentz factors of $\sim10^{6}
\ \Gamma_2\nu_{p,6}^{-1}(1+z)^{-1}$, a similar value to the electrons
produced by $\pi^+$ decay, see eq (\ref{eq:gammae}). If the optical depth
of \(\gamma+\gamma\) pair production is much less than one, then the
photons will escape the GRB jet, and the $\pi^0$ will not affect the
\(\sim 100\) MeV flux.

The $\pi^+$ decays as $\pi^+\rightarrow \mu^++\nu_\mu$, and the
anti-muon decays further as \(\mu^+\rightarrow
e^++\bar{\nu}_\mu+\nu_e\). Isospin conservation arguments suggest a
branching ratio for the delta resonance for $\pi^+:\pi^0$ of
$1:2$. However, for protons interacting with a power-law distribution
of photons, where there are a sufficient number of high-energy photons
to excite higher energy resonances as well as allow direct pion
production, the ratio of charged pions, $\pi^\pm$, to neutral pions is
actually closer to $2:1$. This ratio is more or less independent of the photon
index \citep{Rachen98}. As an approximation, we take the cross section
of the delta resonance, but only consider the high energy electrons
produced by the $\pi^\pm$ decay. This underestimates the
  high energy electron production rate by a factor of 3 when the GRB jet is opaque to
  photons from the $\pi^0$ decay.

\subsection{Analytical Estimate}\label{pion_est}
 
As with the $\pi^0$, at the threshold energy, the $\pi^+$ (and subsequently
 the $\mu^+$) are produced almost at rest in the rest frame of the proton and
therefore have the same Lorentz factor as the proton given in eq
(\ref{eq:min_lf_p}). On average, the positron carries roughly
one-third of the energy of the muon (the remaining two-thirds goes to
neutrinos). The Lorentz factor of the $e^+$ in the comoving jet
rest frame is
\begin{equation}\label{eq:gammae}
\gamma_e\sim\frac{1}{3}\frac{m_\mu}{m_e}\gamma_p\sim10^6\Gamma_2\nu_{p,6}^{-1}(1+z)^{-1},
\end{equation}
\(m_\mu\) and \(m_e\) are the muon and electron masses respectively. 

By requiring that a typical positron produced through pion decay, with
Lorentz factor given by eq (\ref{eq:gammae}) and charge $q$, radiates
at a desired frequency $\nu$ ($\sim 100$ MeV for Fermi-LAT), we solve
for the magnetic field in the jet rest frame, $B$, in Gauss.
\begin{equation}\label{eq:photo-pion_B}
\frac{qB\gamma_e^2\Gamma}{2\pi m_ec(1+z)}\sim 1.6\times
10^{-4}\nu_{8}\  {\rm erg}\Rightarrow
B\sim100\nu_{8}\nu^2_{p,6}(1+z)^3\Gamma_{2}^{-3}\ {\rm Gauss}
\end{equation}
This value for $B$ requires the minimal proton energy to match an
observed flux. If the energy requirements are too large when $B$ is
equal to eq (\ref{eq:photo-pion_B}), they will be even worse when the
magnetic field is not equal to eq
(\ref{eq:photo-pion_B}).\footnote{Equation (\ref{eq:photo-pion_B}) is
  close to the magnetic field value requiring the minimal proton
  energy when the peak frequency of a positron with Lorentz factor
  given by eq (\ref{eq:gammae}) is above the cooling frequency
  $\nu_c$. If a typical photo-pion produced positron is not cooled by
  synchrotron radiation, for typical GRB spectra and efficient proton
  acceleration, the minimum necessary proton luminosity will occur at
  a $B$ such that $\nu_c$ is 100 MeV. For the vast majority of allowed
  GRB parameter space, the photo-pion peak is cooled, so the statement
  that eq (\ref{eq:photo-pion_B}) is a best case scenario holds.}

The observed specific synchrotron flux, \(f_\nu\), is
\begin{equation}
f_\nu=\frac{\sqrt{3}q^3BN_e\Gamma(1+z)}{4\pi d_L^2m_ec^2}
\approx1.2\ \mu {\rm Jy}\ N_{e,50}B\Gamma d_{L,28}^{-2}(1+z),
\end{equation}
where \(N_e\) is the number of electrons that radiate at $\nu$, and
$d_L$ is the luminosity distance to the GRB. Thus, the number of $e^+$
needed to produce the observed flux at $\nu$ is
\begin{equation}\label{eq:ElectronEst}
N_e\approx8\times10^{47}\frac{f_{\nu,\mu{\rm Jy}}d^{2}_{L,28}}{B\Gamma_{2}(1+z)}.
\end{equation}
where \(f_{\nu,\mu{\rm Jy}}\) is the observed flux in $\mu{\rm Jy}$.

The number of protons, with energy above the pion production threshold,
required to produce the necessary electrons in eq
(\ref{eq:ElectronEst}) is \(N_p\approx\frac{3}{2} N_e/\tau_{p\gamma}\), where
$\tau_{p\gamma}$ is the probability that a photon of frequency
$\sim\nu_p(1+z)/\Gamma$ collides with a proton with a Lorentz factor
given by eq (\ref{eq:min_lf_p}) and produces a pion. This probability
is approximately equal to the optical depth to pion production and is
given by \(\tau_{p\gamma }=\sigma_{p\gamma}n_\gamma
R/\Gamma\), where $\sigma_{p\gamma}$ is the cross section for the delta
resonance, $\sigma_{p\gamma}=5\times10^{-28}\ {\rm cm^2}$. $n_\gamma$
is the number density of photons in the comoving frame of the GRB jet,
and $R$ is the distance from the center of the explosion. For a GRB of
observed isotropic luminosity $L_\gamma$ (integrated over the sub-MeV
part of the Band function spectrum) and observed peak frequency
$\nu_p$ (in eV), the number density of photons in the comoving frame
of a GRB jet is
\begin{equation}\label{nphoton}
n_\gamma=\frac{L_\gamma(1+z)^{-1}}{4\pi R^2\Gamma c\nu_p(1.6\times10^{-12}\ {\rm erg/eV})}
\approx
2\times10^{14}L_{\gamma,52}R_{15}^{-2}\Gamma_{2}^{-1}\nu_{p,6}^{-1}(1+z)^{-1}\ 
{\rm cm}^{-3}.
\end{equation}
This gives an optical depth of 
\begin{equation}\label{eq:optical_depth}
\tau_{p\gamma}\approx 0.8
L_{\gamma,52}R_{15}^{-1}\Gamma_{2}^{-2}\nu_{p,6}^{-1}(1+z)^{-1}.
\end{equation}
Using (\ref{nphoton}) and (\ref{eq:optical_depth}) the 
number of protons needed to produce a specific flux, $f_\nu$, is
\begin{equation}
N_p\approx 10^{48}f_{\nu,\mu{\rm
    Jy}}d^{2}_{L,28}\Gamma_2R_{15}\nu_{p,6}B^{-1}L_{\gamma,52}^{-1}.
\end{equation}
The corresponding energy in these protons is
\begin{equation}\label{eq:energy_est}
E_p\approx N_p(\gamma_p m_p c^2)\Gamma\approx 3.0\times 10^{51}
\ \frac{\Gamma_{2}^3f_{\nu,\mu{\rm Jy}}d_{L,28}^2R_{15}}{B
  L_{\gamma,52}(1+z)} \ {\rm erg}.
\end{equation}
It is more useful to consider the luminosity carried by these protons,
$L_p$, as this can be directly compared to the observed
$\gamma$-ray luminosity, $L_\gamma$. The ratio $L_\gamma/L_p$ will
allow us to determine the efficiency of the photo-pion process for the
generation of \(\ga 100\ {\rm MeV}\ \gamma\)-rays. The proton luminosity
is related to $E_p$ by
\begin{equation}\label{eq:proton_lum}
L_p=E_p\Gamma\times\max{\left\{t_{\rm dyn}^{-1},t_{\rm cool}^{-1}\right\}},
\end{equation}
where $t_{\rm dyn}$ is the dynamical time in the jet comoving frame,
\begin{equation}\label{eq:t_dyn}
t_{\rm dyn}=\frac{R}{2c\Gamma}\approx 170 R_{15}\Gamma_{2}^{-1}\ {\rm s},
\end{equation}
and $t_{\rm cool}=\gamma_e m_e c^2/P_{syn}$ is the synchrotron cooling
time in the jet frame. $P_{\rm syn}$ is the total synchrotron power radiated
by a positron. Using the magnetic field in the jet comoving frame given
in eq (\ref{eq:photo-pion_B}), \(t_{\rm cool}\) becomes
\begin{equation}\label{eq:synch_cool}
t_{\rm cool}=\frac{6\pi m_ec}{\sigma_T  B^2 \gamma_e}\approx
8\times10^{-2}\ 
\frac{\Gamma_{2}^5}{\nu^2_{8}\nu_{p,6}^3(1+z)^5} \ {\rm s}.
\end{equation}
Inverse Compton cooling is neglected because the electrons
considered here have a Lorentz factor of $\ga10^6$; so the IC radiation
is greatly decreased because of Klein-Nishina suppression.

Substituting Equations (\ref{eq:photo-pion_B}), (\ref{eq:energy_est}),
(\ref{eq:t_dyn}), \& (\ref{eq:synch_cool}) into (\ref{eq:proton_lum}),
we find the minimum proton luminosity necessary to match a flux of
$f_{\nu,\mu{\rm Jy}}$ at $\nu_8$ with the photo-pion process is
\begin{equation}\label{eq:photo-pion_L_est}
L_{p}
\ga
\left\{ \begin{array}{ll}\hskip -5pt
2\times10^{49}\ \Gamma_{2}^{8}L^{-1}_{\gamma,52}
\nu^{-1}_{8}\nu^{-2}_{p,6}
f_{\nu,\mu{\rm Jy}}d_{L,28}^2(1+z)^{-4}\ {\rm erg\ s^{-1}}
& t_{\rm dyn}<t_{\rm cool}\\ \\
\hskip -5pt
4\times10^{52}\ \Gamma_{2}^{2}R_{15}L^{-1}_{\gamma,52}\nu_8\nu_{p,6}f_{\nu,\mu{\rm Jy}}
d_{L,28}^2(1+z)\ {\rm erg\ s^{-1}}
& t_{\rm cool}<t_{\rm dyn}.
\end{array} \right.
\end{equation}
At first glance, the proton luminosity does not seem prohibitively
large. However, the strong dependence on $\Gamma$ has the
potential to increase the proton energy requirement tremendously. 

To assess the viability of the photo-pion process producing the
$\ga 100$ MeV photons detected by {\it Fermi}, let us consider a
bright {\it Fermi}-LAT burst, GRB 080916C
\citep{2009Sci...323.1688A}. This burst was detected at a redshift of
4.3, which has a corresponding $d_{L,28}=12$. The peak of the observed
spectrum was at 400 keV, and the flux at 100 MeV during the prompt
emission was $f_\nu\sim3\ \mu{\rm Jy}$. The $\gamma$-ray isotropic
luminosity for GRB 080916C was $L_{\gamma,52}\sim20$, and the minimum
jet Lorentz factor was estimated to be $\Gamma_2\sim9$
\citep{2009Sci...323.1688A}. For these parameters, we find $t_{\rm
  cool}<t_{\rm dyn}$ as long as $R>10^{15}\ {\rm cm}$, implying that
the required luminosity in protons with $\gamma_p\ga10^5$ is
$L_p\sim 1.5\times 10^{56}\ R_{15}\ {\rm erg\ s^{-1}}$. This is a
factor of $\sim 700$ times larger than the $\gamma$-ray luminosity at
$R=10^{15}\ {\rm cm}$. Below $R=10^{15}$, \(t_{\rm dyn}<t_{\rm cool}\)
and the proton luminosity has no $R$ dependence.\footnote{See previous
  footnote about how the luminosity estimate may be too pessimistic
  when $t_{\rm dyn}<t_{\rm cool}$. A more accurate calculation that
  accurately minimizes $L_p$ with respect to $B$ for a given $R$,
  $\Gamma$, {\it etc}, is presented in the \S \ref{pion_better}.}  This
efficiency is below the efficiencies of the order 10-20\% estimated
for other GRBs using afterglow modeling
\citep[\textit{e.g.}][]{PK02,2006MNRAS.369..197F,2007ApJ...655..989Z}
and makes the photo-pion process an unlikely candidate to produce the
prompt-LAT emission observed in this GRB.

\subsubsection{Neutrino Flux}\label{pion_nu}
In addition to producing high energy electrons, photo-pion production
also results in high energy neutrinos. For the photo-pion production
models of the observed LAT emission, it is possible to directly
correlate the flux at 100 MeV to an expected flux of neutrinos. Since
there are two muon neutrinos created for every $e^+$ in $\pi^+$ decay, we can
simply use eq (\ref{eq:ElectronEst}) and (\ref{eq:photo-pion_B}) to find
corresponding number of neutrinos,
\begin{equation}\label{eq:nu_number}
N_\nu=10^{46}\ f_{\nu,{\rm \mu Jy}}d_{\rm L,28}^2\nu_8^{-1}\nu_{p,6}^{-2}\Gamma_2^2(1+z)^{-4}
\end{equation}
These neutrinos will have the same energy as the electrons on average,
with an observed energy of
\begin{equation}\label{eq:nu_energy}
E_\nu=10^5 \Gamma_2^2\nu_{p,6}^{-1}(1+z)^{-2}\ {\rm GeV}
\end{equation}

This corresponds to an observed flux, $F_\nu$, at
\(10^5\ \Gamma_2^2\nu_{p,6}^{-1}(1+z)^{-2}\ {\rm GeV}\) of
\begin{equation}\label{eq:nu_flux_calc}
F_\nu=E_\nu(1+z) N_\nu\Gamma\times\frac{1}{4\pi d_L^2}\max{\left\{t_{\rm dyn}^{-1},t_{\rm cool}^{-1}\right\}},
\end{equation}

\begin{equation}\label{eq:nu_flux}
F_{\nu}
\approx
\left\{ \begin{array}{ll}\hskip -5pt
5\times 10^{-7}\ f_{\nu,\mu{\rm Jy}}R_{15}^{-1}\Gamma_{2}^{6}
\nu^{-1}_{8}\nu^{-3}_{p,6}
(1+z)^{-5}\ {\rm GeV\ cm^{-2} s^{-1}}
& t_{\rm dyn}<t_{\rm cool}\\ \\
\hskip -5pt
10^{-3}\ f_{\nu,\mu{\rm Jy}}\nu_8\ {\rm GeV\ cm^{-2} s^{-1}}
& t_{\rm cool}<t_{\rm dyn}.
\end{array} \right.
\end{equation}
The neutrino flux does not depend on the luminosity of the GRB, as it
is fixed by the flux at $\nu_8$, the peak of the photo-pion
synchrotron emission. When the photo-pion electrons are in the cooled
regime---as is expected for much of the GRB parameter space---the
neutrino flux does not depend on any of the jet parameters. The
neutrino flux depends only on the observed LAT flux at 100 MeV; when
the electrons are in the fast cooling regime, the synchrotron flux
depends directly on the electron energy flux, and the neutrino flux depends
directly on the electron energy flux. Of course, the neutrino energy does
depend on $\Gamma$, as seen from eq (\ref{eq:nu_energy}).

To get a rough idea of an expected neutrino count-rate at IceCube, we
fit the averaged effective area for muon neutrinos at IceCube given in
\citet{2012Natur.484..351A} by \(A\approx 100\ {\rm
  m^2}\times\sqrt{E_\nu/(100\ {\rm TeV})}\). Our fit agrees well with the
averaged effective area of 59-string detector when
\(E_\nu>3\times10^4\ {\rm GeV}\). We expect the following neutrino
counts per second of LAT emission from the photo-pion process:
\begin{equation}\label{eq:nu_count_rate}
\frac{dN_\nu}{dt}
\approx
\left\{ \begin{array}{ll}\hskip -5pt
5\times 10^{-6}\ f_{\nu,\mu{\rm Jy}}R_{15}^{-1}\Gamma_{2}^{5}
\nu^{-1}_{8}\nu^{-2.5}_{p,6}
(1+z)^{-4}\ {\rm counts\ s^{-1}}
& t_{\rm dyn}<t_{\rm cool}\\ \\
\hskip -5pt
10^{-2}\ f_{\nu,\mu{\rm Jy}}\Gamma_2^{-1}\nu_8\nu_{p,6}^{0.5}(1+z)\ {\rm counts\ s^{-1}}
& t_{\rm cool}<t_{\rm dyn}.
\end{array} \right.
\end{equation}
Therefore for a bright GRB detected by the {\it Fermi}-LAT with
$f_{\nu,\mu Jy}\sim2$, $\Gamma_2\sim9$, $z\sim2$, $\nu_{p,6}\sim1$,
and $T_{90}=10\ {\rm s}$, We find that $t_{\rm cool}<t_{\rm dyn}$ if
$R>10^{15}$ cm. If $R>10^{15}$ cm, we expect about 0.07 neutrinos of
energy $9.0\times10^5\ {\rm GeV}$ over the course of the burst. If
$R<10^{15}$ cm, the number of neutrinos is increased by a factor
$R_{15}^{-1}$ and the energy stays $9.0\times10^5\ {\rm GeV}$. And so
if we add up the contributions from all {\it Fermi}-LAT bursts, we
expect IceCube to detect $\sim1$ neutrino.

\subsection{Numerical Calculation}\label{pion_better}
The calculation presented in \S\ref{pion_est} is a rough estimate to
the energy requirement for protons, but it doesn't give any spectral
information about the photo-pion radiation and assumes that all the
protons have the same energy. Furthermore, it doesn't take the finite
width of the delta resonance into account. All of these corrections go
in the direction of decreasing the efficiency of the photo-pion
process. A more rigorous calculation is presented in this subsection.

The distribution of photons in the jet rest frame is assumed to be
isotropic, and the Band function is approximated by
\begin{equation}
\frac{dn}{d\epsilon}(\epsilon)=n_{\epsilon_p}
\times
\left\{\begin{array}{ll}\hskip -5pt
\left(\frac{\epsilon}{\epsilon_p}\right)^{-\alpha} 
& \mbox{for } \epsilon_{\rm min}\leq\epsilon\leq \epsilon_p\\ \\
\hskip -5pt
\left(\frac{\epsilon}{\epsilon_p}\right)^{-\beta} 
& \mbox{for } \epsilon_p<\epsilon.
\end{array}\right.
\end{equation}
Here \(\epsilon_p=h\nu_p(1+z)/(\Gamma m_ec^2) \approx.02\Gamma_2^{-1}
\nu_{p,6}(1+z)\) is the dimensionless photon energy at the peak of the
spectrum in the jet comoving frame, and $\epsilon_{\rm min}$ is the
dimensionless photon energy below which the Band function no longer
fits the observed GRB spectrum. $\epsilon_{\rm min}$ is poorly
constrained by GRB observations, and depends on the prompt radiation
mechanism and radius of emission for the prompt emission. If
$\epsilon_{\rm min}$ corresponds to the synchrotron self-absorption
frequency, it is likely of order $10^{-7}$, corresponding to an
observed frequency of a few eV. We more conservatively assign an
$\epsilon_{\rm min}$ value based on the lowest observed frequencies by
{\it Fermi} $\nu_{\rm min}\sim 1\ {\rm keV}$, with a corresponding
$\epsilon_{\rm min} \approx 2\times10^{-5}\Gamma_2^{-1}\nu_{\rm
  min,3}(1+z)$. In reality, the choice of $\epsilon_{\rm min}$ has very
little effect on the synchrotron flux in the {\it Fermi}-LAT band, as
the photo-pion positrons and electrons produced from a proton
interacting with photons of energy $\epsilon_{\rm min}$ will be of
very high Lorentz factor, $\gamma_e\sim 10^9$. For our
calculation, $\epsilon_{\rm min}$ corresponds to a $\nu_{\rm min}$ of
$\sim 1\ {\rm keV}$ and is included in our calculations only for
completeness.

If $\alpha\sim 1$ and the GRB has an isotropic luminosity
$L_{\gamma}$, then the value for the number density of photons per
$m_e c^2$ at \(\epsilon_p\), $n_{\epsilon_p}$, is
\begin{equation}
n_{\epsilon_p}=8\times 10^{15}L_{\gamma,52} R_{15}^{-2}
{\nu_{p,6}}^{-2}\left(1+z\right)^{-2} \quad
{\rm cm^{-3}}.
\end{equation}
We assume that the proton number distribution is a power law,
\begin{equation}
dN_p(\gamma_p)
=N_{p,i}\left(\frac{\gamma_p}{\gamma_i}\right)^{-p}{d\gamma_p}\quad\gamma_i<\gamma_p<\gamma_{\rm max}.
\end{equation}
with a minimum Lorentz factor $\gamma_i\sim 10$ and a maximum Lorentz
factor given by requiring the protons to be confined to the jet, {\it
  i.e.} the Hillas criterion, \(\gamma_{\rm max}=\frac{qBR}{\Gamma
  m_pc^2} =3\times10^6BR_{15}\Gamma_2^{-1}\)
\citep{Hillas84}. $N_{p,i}$ is the number of protons in the emitting
region of a GRB between $\gamma_i$ and $\gamma_i+d\gamma_p$. As these
high energy protons travel through the jet, they will interact with
the photons that make up the Band function, creating secondary
particles. The total interaction rate, $\dot{N}_{p\gamma}$, for a
proton with Lorentz factor $\gamma_p$ and a photon with energy
$\epsilon$ depends on the angle-integrated cross section:
\(\sigma_{p\gamma}(\epsilon')\).  $\epsilon'$ is the energy of the
photon in the nuclear rest frame, $\epsilon' = \gamma_p \epsilon
(1-\beta_p\mu)$, where $\mu$ is the cosine of the angle between the
proton and photon and $\beta_p$ is the velocity of the proton divided by
$c$.  The interaction rate is
\begin{equation}
\dot{N}_{p\gamma}=\frac{c}{4\pi}
\int{d\Omega
  \int{d\epsilon\ n(\epsilon,\Omega)(1-\beta_p\mu)
    \sigma_{p\gamma}(\epsilon')}}.
\end{equation}
Since $\gamma_p\gg 1$, $\beta_p\sim 1$, and we are approximating the
sub-MeV Band photons as isotropic in the rest frame of the jet, the
number of scatterings is approximated by
\begin{equation}\label{eq:scatter_rate}
\frac{d\dot{N}_{p\gamma}}{d\gamma_p}=\frac{c}{2\gamma_p^2}\frac{dN_p}{d\gamma_p}
\int_0^\infty{d\epsilon\ \frac{n(\epsilon)}{\epsilon^2}
  \int_0^{2\gamma_p\epsilon}{d\epsilon'\ \epsilon'
\sigma_{p\gamma}(\epsilon')}
  }.
\end{equation}
We approximate the cross section of the delta resonance,
$\sigma_{p\pi}(\epsilon')$, as $5\times10^{-28}\ {\rm cm^2}$
if $530<\epsilon'<760$ and 0 otherwise. As before, we treat the
pion and muon as decaying instantaneously without any energy losses
and approximate \(\gamma_e = 70\gamma_p\). Equation
(\ref{eq:scatter_rate}), re-written in terms of the produced
electrons, is
\begin{equation}\label{eq:pion_production}
\frac{d\dot{N}_e}{d\gamma_e}=
\frac{c}{10^4\gamma_e^2}\frac{dN_p}{d\gamma_p}\frac{d\gamma_p}{d\gamma_e}
\int_0^\infty{d\epsilon\ \frac{n(\epsilon)}{\epsilon^2}
  \int_0^{2\gamma_p\epsilon}{d\epsilon'\ \epsilon'
\sigma_{p\gamma}(\epsilon')}}.
\end{equation}
When evaluating the scattering rate, it is convenient to define two
electron Lorentz factors of interest: $\gamma_{\rm peak}$, the
electron that is produced from a proton interacting with photons at
the observed peak in the gamma-rays, and $\gamma_{\rm break}$, the
electron that is produced from a proton interacting with the lowest
energy photon in the Band function, which is taken to be $\nu_{\rm
  min}$. $\gamma_{\rm peak}$ and $\gamma_{\rm break}$ are equal to
\begin{eqnarray}
\gamma_{\rm peak}&=&1.2\times10^6\Gamma_{2}\nu_{p,6}^{-1}(1+z)^{-1}\label{eq:lf_peak}\\
\gamma_{\rm break}&=&1.2\times10^9\Gamma_{2}\nu_{\rm min,3}^{-1}(1+z)^{-1}.
\end{eqnarray}

Carrying out the integration in equation (\ref{eq:pion_production}),
we find the rate of electrons produced through the photo-pion processes is
\begin{eqnarray}
\frac{d\dot{N}}{d\gamma_e}&\approx&
\left\{ \begin{array}{ll}\hskip -5pt
\displaystyle N_{e,p}\left(\frac{\gamma_e}{70\gamma_i}\right)^{\beta-p-1}
& 70\gamma_{i}\leq\gamma_e\leq \gamma_{\rm peak}\\ \\
\hskip -5pt
\displaystyle N_{e,p}\left(\frac{\gamma_{\rm peak}}{70\gamma_i}\right)^{\beta-p-1}\left(\frac{\gamma_e}{\gamma_{\rm peak}}\right)^{\alpha-p-1}
& \gamma_{\rm peak} < \gamma_e\leq \gamma_{\rm break}\\ \\
\hskip -5pt
\displaystyle N_{e,p}
\left(\frac{\gamma_{\rm peak}}{70\gamma_i}\right)^{\beta-p-1}\left(\frac{\gamma_{\rm break}}{\gamma_{\rm peak}}\right)^{\alpha-p-1}\left(\frac{\gamma_e}{\gamma_{\rm break}}\right)^{-p-2}
& \gamma_{\rm break}<\gamma_e
\end{array} \right.\\ \\
N_{e,p}&\approx&0.3(6\times10^{-5})^\beta N_{p,i}\gamma_i^{\beta-1} L_{\gamma,52}
R_{15}^{-2}\nu_{p,6}^{\beta-2}\left(1+z\right)^{\beta-2}\Gamma_2^{-\beta}/(\beta+1).
\end{eqnarray}
We now derive the previous estimate of how much energy the
protons would need to carry to produce the observed Fermi-LAT flux at
100 MeV.

For simplicity, we set $\alpha$ and $\beta$ to the typical GRB
parameters $\alpha=1$ and $\beta=2.2$. We assume the protons have a power
law index of $p=2$ and $\gamma_i=10$, corresponding to efficient acceleration in
shocks. Then, the majority of the energy in the photo-pion electrons is
contained in the electrons with $\gamma_{\rm peak}\leq
\gamma_e\leq\gamma_{\rm break}$. This section of the power law is

\begin{equation}
\frac{d\dot{N}}{d\gamma_e}
=7\times10^{-12}N_{p,i}L_{\gamma,52}R_{15}^{-2}
\nu_{p,6}^{-1}(1+z)^{-1}\Gamma_{2}^{-1}
\left(\frac{\gamma_e}{10^6}\right)^{-2}.
\end{equation}
To calculate the total number of electrons produced, we solve the
following continuity equation
\begin{equation}\label{eq:cont_equation}
\frac{\partial N(\gamma_e)}{\partial t}+
\frac{\partial}{\partial
  \gamma_e}\left\{\dot{\gamma}_eN(\gamma_e)\right\}
=\frac{d\dot{N}}{d\gamma_e}.
\end{equation}
We approximately solve this equation by following the standard
procedure of breaking up the continuity equation into two regimes:
one where the electrons are cooling slowly, {\it i.e.}  $t_{\rm
  dyn}<t_{\rm cool}$, and another where cooling losses are important,
$t_{\rm dyn}>t_{\rm cool}$. The solution of differential equation
(\ref{eq:cont_equation}) is then approximately:
\begin{equation}
N(\gamma_e)=
\left\{ \begin{array}{ll}\hskip -5pt
\displaystyle 
t_{\rm dyn}\frac{d\dot{N}}{d\gamma_e}
& t_{\rm dyn}<t_{\rm cool}\\ \\ 
\hskip -5pt
\displaystyle
 \frac{1}{\dot{\gamma}_e}\int_{\gamma_e}^{\infty}{d\gamma \frac{d\dot{N}}{d\gamma}}
& t_{\rm dyn}>t_{\rm cool},
\end{array} \right.
\end{equation}
Two different cooling mechanisms are considered: synchrotron 
and inverse Compton cooling. To see which is the more dominant cooling
process, we compare the power radiated by each process.  The
synchrotron power for electrons with $\gamma_e=\gamma_{\rm peak}$ is
\begin{equation}\label{eq:synch_power}
P_{\rm syn}(\gamma_{\rm peak})=1.6\times10^{-3}\Gamma_{2}^{2}(1+z)^{-2}\nu_{p,6}^{-2}
B^2\ {\rm erg/s}.
\end{equation}
From the condition \(t_{\rm cool}<t_{\rm dyn}\), the electrons at the
peak of the distribution will be cooled via synchrotron if
\begin{equation}
B> 2\ R_{15}^{-0.5}\nu_{p,6}^{0.5}(1+z)^{0.5}\ {\rm Gauss}.
\end{equation}
This is a low value of the magnetic field, so synchrotron cooling
losses are important to consider. However, for completeness, our
estimate will consider both possibilities, when $\gamma_{\rm peak}$ is
above cooling and below cooling.

For inverse Compton losses, while the energy density in the photons
can be very large, particularly at distances less than $\sim10^{16}$
cm, the inverse Compton radiated power is greatly reduced due to
Klein-Nishina suppression. For the electron Lorentz factor given in eq
(\ref{eq:lf_peak}), all of the prompt sub-MeV emission will be in the
Klein-Nishina regime if \(\gamma_{\rm peak}\epsilon_{\rm min}>1\), or
\(\nu_{p,6} < 20\nu_{\rm min,3}\). The power radiated due to IC
scattering in the Klein-Nishina regime is given in
\citet{PhysRevD.3.2308}:
\begin{equation}
P_{KN}(\gamma)=m_ec^3\pi r_0^2
\int_{\frac{1}{\gamma}}^\infty{d\epsilon\ \frac{1}{\epsilon}\frac{dn}{d\epsilon}
  \left(\log{(4\gamma\epsilon)}-\frac{11}{6}\right)}.
\end{equation}
where $r_0$ is the classical electron radius. For \(\gamma=\gamma_{\rm
  peak}\), neglecting the logarithmic dependencies of variables and
assuming $\alpha\sim1$, the IC radiated power is
\begin{equation}\label{eq:IC_power}
P_{KN}(\gamma_{\rm peak})=2\times10^{-1}L_{\gamma,52}R_{15}^{-2}
\nu_{p,6}^{-2}(1+z)^{-2}
\left(\frac{\nu_{p,6}}{\nu_{\rm min,3}}\right).
\end{equation}
From equations (\ref{eq:synch_power}) and (\ref{eq:IC_power}), the
ratio of synchrotron power to Inverse Compton power at $\gamma_{\rm
  peak}$ is
\begin{equation}
\frac{P_{syn}}{P_{KN}}=8\times10^{-3}L_{\gamma,52}^{-1}\Gamma_{2}^2 B^2 R_{15}^2
\left(\frac{\nu_{\rm min,3}}{\nu_{p,6}}\right).
\end{equation}
If we define \(\epsilon_B\) as the ratio of energy density in the
magnetic field to energy density in radiation, the ratio becomes
\begin{equation}\label{eq:pow_ratio}
\frac{P_{syn}}{P_{KN}}=50\epsilon_{B,-2}\left(\frac{\nu_{\rm min,3}}{\nu_{p,6}}\right).
\end{equation}
Unless $\epsilon_B$ is small, the synchrotron emission will dominate
over the inverse Compton emission. The inverse Compton scattered
photons will have on average an energy in the jet's rest frame of
$\gamma_e m_ec^2\sim 0.5\ {\rm TeV}$ for electrons with
$\gamma_e=\gamma_{\rm peak}$. These photons will quickly pair produce
and form a cascade of secondary particles. A full treatment of this is
beyond the scope of this paper. In any case, as can be seen in
equation (\ref{eq:pow_ratio}), the energy in this cascade will be less
than the synchrotron energy radiated by the photo-pion produced
electrons; this allows us to ignore these $e^\pm$ pairs when for
estimating the flux at 100 MeV.

The electron number distribution created by the photo-pion process
for \(\gamma_{\rm peak}\leq\gamma_e\leq\gamma_{\rm break}\) is
\begin{equation}
\frac{dN}{d\gamma_e}=
\left\{ \begin{array}{ll}\hskip -5pt
10^{-9}\ N_{p,i}L_{\gamma,52}R_{15}^{-1}\nu_{p,6}^{-1}(1+z)^{-1}\Gamma_{2}^{-2}\gamma_{e,6}^{-2}
& t_{\rm dyn}<t_{\rm cool}\\ \\
\hskip -5pt
5\times10^{-9}\ B^{-2}N_{p,i}L_{\gamma,52}R_{15}^{-2}\nu_{p,6}^{-1}(1+z)^{-1}\Gamma_{2}^{-1}\gamma_{e,6}^{-3}
& t_{\rm dyn}>t_{\rm cool}.
\end{array} \right.
\end{equation}
The observed synchrotron flux, $f_v$, at $\nu\sim 100\ {\rm MeV}$, is
calculated using the following approximation for synchrotron
radiation:
\begin{eqnarray}
f_{\nu}&=&(1+z)\int_{\gamma_\nu}^{\gamma_{\rm max}}{d\gamma_e
  \frac{\sqrt{3}q^3\Gamma B
 N(\gamma_e)}{4\pi d_L^2m_ec^2}
\left(\frac{\gamma_\nu}{\gamma_e}\right)^{2/3}},\\
\gamma_\nu^2&=&\frac{2\pi m_ec(1+z)\nu}{qB\Gamma (4.13\times10^{-15}\ {\rm eV\ s})}
.\end{eqnarray}
Using equation (\ref{eq:photo-pion_B}) for the magnetic field value,
we  calculate the necessary luminosity in protons to produce
$\gamma$-rays through the photo-pion process. The result we found for $L_p$ is 
\begin{equation}\label{eq:better_L_est}
L_{p}=
\left\{ \begin{array}{ll}\hskip -5pt
3\times10^{51}\ \Gamma_{2}^{8}L^{-1}_{\gamma,52}\nu_8^{-1}\nu_{p,6}^{-2}f_{\nu,\mu{\rm Jy}}d^2_{L,28}(1+z)^{-4}\ {\rm erg\ s^{-1}}
& t_{\rm dyn}<t_{\rm cool}\\ \\
\hskip -5pt
7\times10^{54}\ \Gamma_{2}^{2}R_{15}L^{-1}_{\gamma,52}\nu_8\nu_{p,6}f_{\nu,\mu{\rm Jy}}d^2_{L,28}(1+z)\ {\rm erg\ s^{-1}}
& t_{\rm dyn}>t_{\rm cool}.
\end{array} \right.
\end{equation}
In comparison to the previous estimate for $L_p$ given in equation
(\ref{eq:photo-pion_L_est}), the values for $L_p$ in equation
(\ref{eq:better_L_est}) are considerably larger, by a factor of
$\sim100$. A factor of $\sim 20$ is attributable to the fact that
unlike the estimate given in \S \ref{pion_est}, this calculation
considers protons that are part of a power law distribution that
extends over several decades of energy. Additional factors come from
the finite width of the delta resonance and keeping track of the
factors that come from integration. For the parameters of GRB 080916C,
for the expected case $t_{\rm cool}<t_{\rm dyn}$, the required
proton luminosity is $L_p\sim 3\times 10^{58}\ R_{15}\ {\rm
  erg\ s^{-1}}$. So, the luminosity in the protons is $10^{5}$ times
larger than luminosity in the \(\gamma\)-rays at $R_{15}$, which is
too large to be realistic for a stellar mass object.

\begin{figure*}
\begin{center}\includegraphics[width=.9\textwidth]{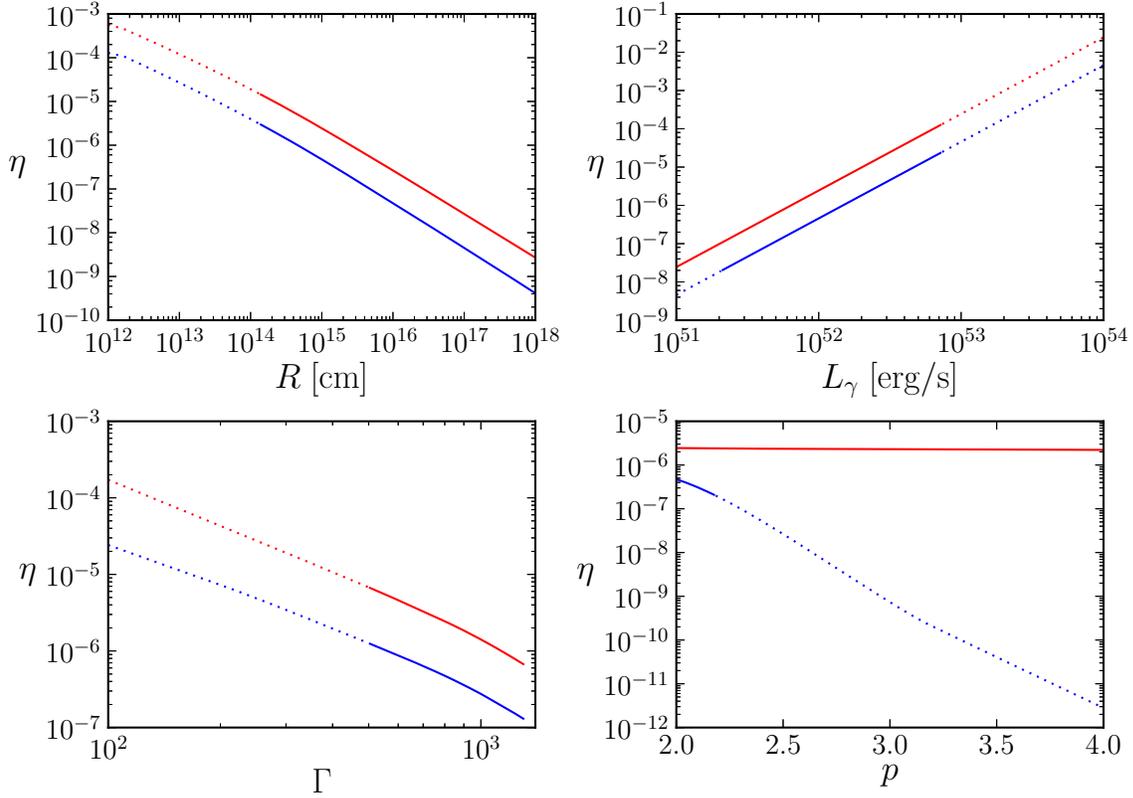}\end{center}

  \caption{A plot of the maximum efficiencies for the photo-pion
    process radiating a typical {\it Fermi}-LAT flux of 1 $\mu$Jy at
    100 MeV as a function of $R$, $L$, $\Gamma,$ and $p$. The lower
    blue lines corresponds to a more physically realistic case of a
    proton power law extending from a Lorentz factor of 10 to the
    Hillas criterion for the confinement of protons. The upper red
    lines are the efficiency only considering the energy in the
    protons that produce the pions and then electrons that radiate at
    LAT frequencies. These red lines represent an absolute maximum
    possible efficiency and corresponds roughly to our calculation in
    \S \ref{pion_est}. The dotted line corresponds to the cases when
    LAT emission could not be seen by an observer, either because the
    emission happens below the photosphere or because the jet would be
    opaque to photons of 10 GeV due to $\gamma + \gamma$ pair
    production. In a two zone model for gamma-ray
      generation, the area where this occurs may differ by a small
      factor. When the parameters are not on the $x$-axis,
    values of $L_\gamma=10^{52}$ erg/s, $R=10^{15}$ cm, $\Gamma=800$,
    \(\nu_p=1\) MeV, $z=2$, $d_L=4.9\times10^{28}$ cm, and $p=2$ are
    taken. The photon power law indices of the Band function were set
    to $\alpha=1$ and $\beta=2.2$. Since $\nu_p,$ $\alpha,$ and
    $\beta$ are unable to change the maximum $\eta$ by more than an
    order of magnitude, their corresponding plots are omitted.}
\label{fig:efficiency}
\end{figure*}

We define the efficiency, $\eta$, as
\begin{equation}\label{eq:eta}
\eta\equiv \frac{L_{\gamma}}{L_p}=
\left\{ \begin{array}{ll}\hskip -5pt
3\ \Gamma_{2}^{-8}L^{2}_{\gamma,52}\nu_8\nu_{p,6}^{2}f^{-1}_{\nu,\mu{\rm Jy}}d^{-2}_{L,28}(1+z)^4
 &  t_{\rm dyn}<t_{\rm cool}\\ \\
\hskip -5pt
10^{-3}\ \Gamma_{2}^{-2}R^{-1}_{15}L^{2}_{\gamma,52}\nu^{-1}_8\nu^{-1}_{p,6}f^{-1}_{\nu,\mu{\rm Jy}}d^{-2}_{L,28}(1+z)^{-1}
& t_{\rm dyn}>t_{\rm cool}.
\end{array} \right.
\end{equation}
In the previous equation, the cooled and uncooled estimates for $\eta$
were calculated by choosing a magnetic field to ensure the energy peak
of the photo-pion-produced electrons radiated at 100 MeV. While this
is convenient and pretty accurate maximum efficiency for analytical
estimation, we also numerically calculated the maximum efficiency,
allowing $B$ to be a free parameter while fixing all the other
parameters ($R$, $\Gamma$, $L_\gamma$, $\nu_p$, {\it etc}). As bounds
on $B$, we set the minimum magnetic field value by requiring that the
power radiated through inverse Compton is no more than 100 times the
synchrotron power for an electron that has a synchrotron peak at 100
MeV. We set a maximum value for $B$ such that the energy in the
magnetic field is at most 10 times the energy in the photons. For the
parameter space we considered, the $B$ that maximized $\eta$ was well
within these bounds.  For a given $L$, $\Gamma$, $R$, and $p$, we
calculate the maximum efficiency of photo-pion electrons radiating the
desired flux of 1 $\mu$Jy at 100 MeV. This maximum efficiency is
plotted in figure \ref{fig:efficiency}. The part of equation (\ref{eq:eta}) corresponding
to fast electron cooling gives an accurate prediction of the maximum
$\eta$. In the slow cooling regime, equation (\ref{eq:eta}) predicts
too small a value of $\eta$; in this case the maximum efficiency is
found when $B$ is a value such that 100 MeV is $\nu_c$.

\begin{figure*}
\begin{center}\includegraphics[width=.9\textwidth]{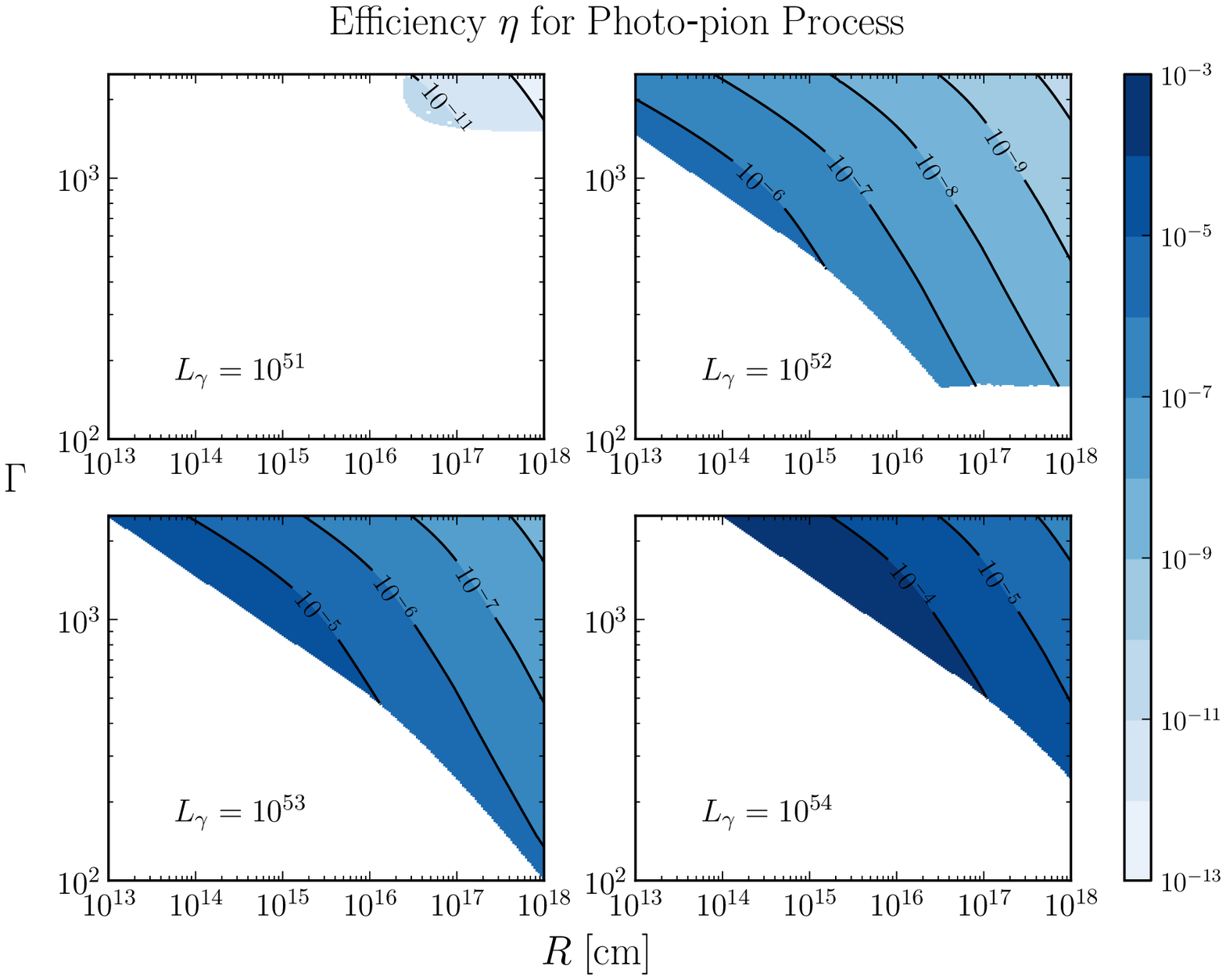}\end{center}
\caption{The maximum $\eta$ is plotted in the $R-\Gamma$ plane for
  various values of $L_\gamma$ to match a flux of $f_\nu=1\ \mu{\rm
    Jy}$ at 100 MeV. As in Figure \ref{fig:efficiency}, we fix
  \(\nu_p=1\) MeV, $z=2$, $d_L=4.9\times10^{28}$ cm, and $p=2$. The
  photon power law indices were set to typical values for the Band
  function, $\alpha=1$ and $\beta=2.2$. Where $\eta$ is not shown, the
  place of emission is either below the photosphere or opaque to
  radiation of 10 GeV due to $\gamma+\gamma$ pair production.}
\label{fig:2D}
\end{figure*}

As illustrated in figure \ref{fig:efficiency} and equation
(\ref{eq:eta}), $R$, $\Gamma$ and $L_\gamma$ are the only parameters
capable of changing $\eta$ significantly; $p$ can as well, but it is
fixed by the desired photo-pion spectrum and therefore not a free
parameter. From typical GRB spectra, we expect $p$ to be in the rage
2.4--2.8 to match typical LAT spectra. In the bottom right panel of
figure \ref{fig:efficiency}, we can see that $p$ has almost no effect
on $\eta$ when we only consider the protons creating the $>100$ MeV
photons (see upper red line in figure \ref{fig:efficiency}). In figure
\ref{fig:2D}, to explore how the efficiency changes with $R$,
$\Gamma$, and $L_\gamma$ we plotted $\eta$ in the $R-\Gamma$ plane for
various $L_\gamma$. It is interesting to note that although $\eta$
scales as $L_\gamma^2$ for a fixed $R,\ \Gamma$, the maximum
efficiency in the $R-\Gamma$ plane only scales as $\sim L_\gamma$
because the available parameter space decreases with increasing
$L_\gamma$ due to $\gamma + \gamma$ pair production.


\section{Bethe-Heitler Pair Production}\label{Bethe_Heitler}
Through Bethe-Heitler pair production, protons and photons interact to
create electron-position pairs directly, $p+\gamma\rightarrow
p+e^++e^-$.  The Bethe-Heitler cross-section and the energy of the
produced electron-positron pair depend strongly on the angle between
the outgoing electron/positron and the proton. Therefore, it is not
possible to use the integrated cross section to calculate the
secondary electron production. Assuming that the protons and photons
are isotropic in the jet's rest frame and using the head on
approximation, {\it i.e.}  the angle between the photon and proton is
zero, {\it i.e.} \(\epsilon'=2\gamma_p\epsilon\), the equation for the rate of
production of secondary electrons is:
\begin{equation}\label{eq:pairpro}
\frac{d\dot{N}_e}{d\gamma_e}(\gamma_e)=2c\int_0^\infty{d\epsilon\ 
    n(\epsilon)\int_1^\infty{d\gamma_p\ N_p(\gamma_p)
      \frac{d\sigma(\epsilon,\gamma_p)}{d\gamma_e}}}.
\end{equation}
In this equation $N_p(\gamma_p)$ is the number of protons with Lorentz
factor $\gamma_p$ and $n(\epsilon)$ is number density of photons with
energy $\epsilon$.  The formula for the differential Bethe-Heitler
cross section, \(\sigma_{BH}\), in the Born approximation, integrated
over angles in the highly relativistic regime, was derived by
\citet{1954PhRv...93..768B} (see \citet{Rachen96} for a more recent
review).
\begin{equation}
\frac{d\sigma_{BH}}{d\gamma'_+}=\frac{3\alpha_f\sigma_T}{2\pi
{\epsilon'}^3}
\left({\gamma_+'}^2+{\gamma_-'}^2
+\frac{2}{3}\gamma_+'\gamma'_-\right)
\left(\log{\frac{2\gamma'_+\gamma'_-}{\epsilon'}}-\frac{1}{2}\right).
\end{equation}
In this equation, $\gamma'_+ (\gamma'_-)$ is the Lorentz factor of the positron
(electron), \(\alpha_f\) is the fine structure constant, and all of
the above quantities are in the proton rest frame. Much of the
contribution to the angle-integrated cross section comes from angles
between the photon and outgoing $e^\pm$ of
order \(\theta'_{\pm}=\frac{1}{\gamma_{\pm}'}\). When
$\gamma_p\gg\gamma'_\pm$, the Lorentz factor of $e^\pm$ in the jet's
rest frame is
\begin{equation}
\gamma_{\pm}=\gamma_p\gamma_\pm'
\left(1-\beta_p\beta'_\pm\cos{\theta'_\pm}\right)\approx
\frac{\gamma_p\gamma_\pm'}{2}\left(\gamma_p^{-2}
+{\gamma_\pm'}^{-2}+{\theta'_\pm}^{-2}\right)
\approx \frac{\gamma_p}{\gamma'_{\pm}}.
\end{equation}
Therefore, most pairs produced via the Bethe-Heitler process have
Lorentz factors (in the jet comoving frame) that are smaller than the
proton that created it.

If $\epsilon'\ll m_p/m_e\sim 10^3$, the nuclear recoil of the
proton can be neglected and the following equality holds:
\begin{equation}\gamma'_++\gamma'_-=\epsilon'.
\end{equation}
For large $\epsilon'$, the differential cross section decreases very
rapidly when $\gamma_\pm'<2$. Therefore, we only consider
$\gamma_\pm'\geq 2$, where the differential cross section is more or
less constant. In this regime, the differential cross section
simplifies to
\begin{equation}\label{eq:diff-cross-NRF}
\frac{d\sigma_{BH}}{d\gamma'_+}\approx\frac{\alpha\sigma_T}{\epsilon'},
\mbox{ if } 2\leq\gamma_+'\leq\epsilon'-2.
\end{equation}
Re-writing eq (\ref{eq:diff-cross-NRF}) in the jet
comoving frame and using the $\epsilon'\approx 2\gamma_p\epsilon$,
we find
\begin{equation}\label{eq:diff-cross-jet}
\frac{d\sigma_{BH}}{d\gamma_+}\approx\frac{\alpha\sigma_T}{2\epsilon\gamma_+^2},
\mbox{ if } \frac{1}{2\epsilon}\leq\gamma_+\leq \frac{\gamma_p}{2}.
\end{equation}

The integral in equation (\ref{eq:pairpro}) is simplified by
considering this approximate expression for the cross-section. The
integral is now straight forward to calculate for a Band spectrum with
indices $\alpha$, $\beta$ and a proton index of $p$. The result is

\begin{equation}\label{eq:pairpro_general}
\frac{d\dot{N}_e}{d\gamma_e}(\gamma_e)\approx
\left\{
\begin{array}{ll}\hskip -5pt
\frac{2c\alpha_f\sigma_T}{\beta(p+1)\gamma_e}n_{\epsilon_p}N_{p,i}
\left(\frac{\gamma_e \epsilon_p}{5}\right)^\beta
\left(\frac{2\gamma_e}{\gamma_i}\right)^{-p} & \mbox{for }
\frac{\gamma_i}{2}\leq\gamma_e\leq 5/\epsilon_p \\ \\
\hskip -5pt
\frac{2c\alpha_f\sigma_T\epsilon_p}{5\beta(p+1)}n_{\epsilon_p}
N_{p,i}\left(\frac{10}{\epsilon_p\gamma_i}\right)^{-p}
\left(\frac{\epsilon_p\gamma_e}{5}\right)^{\alpha-p-1} & \mbox{for }
5/\epsilon_p \leq\gamma_e\leq 5/\epsilon_{\rm min}.
\end{array}\right.
\end{equation}

We now compare Bethe-Heitler pair production to the photo-pion process. The
integrated cross section for Bethe-Heitler process is roughly 10 times
larger than the cross section for the photo-pion
$\Delta$-resonance. For any given proton Lorentz factor $\gamma_p$,
the photon energy required for Bethe-Heitler is roughly 50 times
smaller than for the $\Delta$-resonance. For a given $\gamma_p$, the
photo-pair will have an average Lorentz factor of $\sim\gamma_p/5$
while the delta resonance will decay to a electron with an energy
$\sim 70\gamma_p$. Consider the case where protons with a power-law
distribution function with index $p$ are scattering with a isotropic
photon power-law spectrum $n(\epsilon)\propto \epsilon^{-a}$. The
ratio of the number of $e^\pm$ above a fixed Lorentz factor generated
by Bethe-Heitler process compared to those generated by photo-pion
process is $\sim 2\times 10\times (10^4)^{a-1} \times
(300)^{-p+1}$---the first factor comes from the fact that
Bethe-Heitler produces a electron-positron pair compared to a single
positron produced in the delta-resonance, the second factor is the
ratio of the total cross sections for the Bethe-Heitler and photo-pion
scatterings, the third factor accounts for the larger number of
photons that participate in the Bethe-Heitler process (the threshold
energy for Bethe-Heitler is $\epsilon\sim \gamma_e^{-1}$ and the
threshold energy for photo-pion is $\sim10^{4}\gamma_e^{-1}$) and the
final factor is due to the fewer number of protons that can create
electrons with Lorentz factor $\ga\gamma_e$. This means that which process
dominates depends strongly on which part of the Band function the
protons are interacting with to produce $e^\pm$ with Lorentz factor
$\ga\gamma_e$. For $\gamma_e\ga10^6$, the energy threshold for
both processes lies below the peak of the Band function, so
$a=\alpha\approx 1$. Thus, in this regime, the photo-pion pairs
dominate.

However, for $\gamma_e\la10^3$, the threshold photon energy for
both process is above the peak of the gamma ray spectrum, so
$a=\beta\approx2.2$ and the Bethe-Heitler process is a lot more
efficient than the photo-pion process. Although relativistic shocks
are likely capable of accelerating electrons to $\gamma_e\sim10^3$,
Bethe-Heitler process could still be important if the number of
electrons produced above the photosphere vastly outnumber the
electrons expected to be in the GRB jet from simple charge
neutrality. If the GRB has proton luminosity $L_p$ given by
$L_p=\eta^{-1}L_\gamma$, the comoving electron density is
\(n_e=n_p\approx 2 \times 10^9 \eta^{-1} L_{\gamma,52} \Gamma_2^{-2}
R_{15}^{-2}\ {\rm cm^{-3}}\). The number density of Bethe-Heitler produced electrons,
\(n_{BH}\) is
\begin{equation}
n_{BH} \sim \alpha_f\sigma_Tn_\gamma n_p R/\Gamma \quad \Rightarrow
\quad \frac{n_{BH}}{n_e}=\alpha_f\sigma_Tn_\gamma
R/\Gamma. \end{equation} Since we want to restrict ourselves to above
the photosphere, the optical depth \(\sigma_Tn_e R/\Gamma<1\) or
\begin{equation}
\frac{n_{BH}}{n_e}<\alpha_f\frac{n_\gamma}{n_e}\sim
10^3\eta\Gamma_2\nu_{p,6}^{-1}(1+z)^{-1},
\end{equation}
where $n_\gamma$ is given by eq (\ref{nphoton}). It is somewhat
counter-intuitive, but the Bethe-Heitler process is likely to be most
important in jets with lower baryon loading, {\it i.e.} when $\eta$ is
large. The Bethe-Heitler process could be important for $\gamma_e'\ll
10^5$---especially if for some reason the Fermi mechanism is unable
to accelerate electrons to this Lorentz factor in GRB shocks---but for these
$e^\pm$ to account for the 100 MeV photons from GRBs via the
synchrotron process requires a very large magnetic field and the
luminosity carried by such a magnetic field would greatly exceed
10$^{52}$ erg/s. Therefore, it seems that at best there might just be
a small part of the parameter space for GRBs where the Bethe-Heitler
mechanism could play an interesting role in the generation of prompt
$\gamma$-ray radiation.

\section{Proton Synchrotron}\label{pro_synch}
Massive particles have lower radiative losses than lighter
particles, and therefore more massive particles are easier to
accelerate in shocks. The maximum Lorentz factor that protons can
attain is much larger than the maximum Lorentz factor of
electrons. The maximum synchrotron photon energy from a
shock-accelerated particle is given by requiring that the synchrotron
energy radiated during one acceleration time (on the order of the
Larmor time) is equal the energy gained in an acceleration
cycle---half of the particle's energy. {\it i.e.}  \(\gamma
mc/(qB)\times 4B^2q^4\gamma^2/(9m^2c^3)\la\gamma mc/2\). The maximum
photon energy for a source moving with a Lorentz factor $\Gamma$ at
redshift $z$ is
\begin{equation}\label{eq:max_synch}
\nu_{\rm max}\approx \frac{9\Gamma mc^3}{16\pi
  q^2(1+z)}\sim50\frac{\Gamma}{1+z} \left(\frac{m}{m_e}\right)\ {\rm MeV}.
\end{equation}
While electron synchrotron radiation can only produce photons up to an
energy $\sim 50\Gamma/(1+z)$ MeV, the proton synchrotron process can
radiate photons up to $10^2\Gamma/(1+z)$ GeV. For this reason, when
photons of energies larger than what is allowed by electron
synchrotron are detected from a source, proton synchrotron is
frequently suggested as a possible radiation mechanism
\citep[\textit{e.g.}][]{1998ApJ...499L.131B, 1998ApJ...509L..81T,
  2000NewA....5..377A, 2001ApJ...559..110Z, 2003APh....18..593M,
  2004NewAR..48..411R, Razzaque10}.

However, while the lower radiative efficiency of protons allows the
protons to radiate at higher frequencies, it also means that the
proton-synchrotron model requires more energy in the magnetic field to
match an observed flux. Because of this, we find that to match the typical
observations of {\it Fermi}-LAT GRBs, either the energy requirements
are prohibitive or the proton power-law distribution would have to
begin at extremely high Lorentz factors.

As before, we are considering protons with a power-law distribution
\(dN_p(\gamma_p)\propto\gamma_p^{-p}\ d\gamma_p\) if
\(\gamma_p\geq\gamma_i\). The proton injection frequency, $\nu_i$, is
\begin{equation}\label{eq:pro_i}
\nu_i=\frac{qB\Gamma\gamma_i^2}{2\pi m_pc(1+z)}\approx
6.3\times10^{-10} B\Gamma_2\gamma_i^2(1+z)^{-1}\ {\rm eV}.
\end{equation}
We define the cooling frequency, $\nu_c$, as the frequency where the
synchrotron cooling time of the protons that radiate at $\nu_c$ is
equal to the dynamical time. The cooling time for protons is increased
by a factor $\left(\frac{m_p}{m_e}\right)^3$ compared to the cooling
time of electrons. The cooling frequency for proton synchrotron is
\begin{equation}\label{eq:pro_cool}
\nu_c\approx 5\times10^{23}
B^{-3}R_{15}^{-2}\Gamma^3_2(1+z)^{-1}\ {\rm eV}.
\end{equation}
Since nearly all GRB observed in the {\it Fermi}-LAT band have a
spectrum that can be fit by a single power law in the LAT band, we examine two
possible spectral orderings: $\nu_i\sim\nu_c\leq\nu$ and the slow
cooling regime $\nu_i\leq\nu\leq\nu_c$.

$\nu$ must be above $\nu_i$ to match the spectra of {\it Fermi}-LAT
GRBs: $\nu\leq\nu_i$ cannot produce GRB LAT emission because if the
protons are cooled (uncooled), the spectrum is $f_\nu\propto
\nu^{-1/2}\ (\nu^{1/3})$. These spectra are harder than what is
observed for most GRB, which have a typical high energy $f_\nu$ index
$<-1$ \citep{Fermi:High_Energy}. Therefore, we take $\nu_i\leq\nu$ to
agree with a typical GRB spectrum. The synchrotron flux $f_\nu$ at the
peak of the spectrum (\(\nu_i=\min{(\nu_i,\nu_c)}\)) is
\begin{equation}\label{eq:pro_flux}
f_\nu\approx 7 
BN_{52}\Gamma_2(1+z)d^{-2}_{L,28}\ \mu{\rm Jy},
\end{equation}
where N is the total number of protons radiating in a dynamical
time. The flux scales as $f_\nu\propto\nu^{-\frac{p-1}{2}}$ if
$\nu_i\leq\nu\leq\nu_c$ and as $f_\nu\propto\nu^{-\frac{p}{2}}$ if
$\nu\geq\nu_c,\nu_i$.  

Below $\nu_p$, the flux of a typical GRB is constant, $f_\nu\propto
\nu^0$. Above $\nu_p$, the flux scales as $f_\nu\propto
\nu^{-1.2}$. Since this break is larger than one half, it cannot be
attributed to a cooling break. Therefore, in order to have
$f_\nu\propto\nu^0$ below $\nu_p$, we require that both $\nu_i$ and
$\nu_c$ lie above $\nu_p$. Furthermore, the majority of LAT GRBs show
a single power law above their peak, extending up to a maximum
observed frequency, $\nu_{\rm max}$, on the order of tens of GeV. We
need to ensure that the proton synchrotron radiation does not add any
spectral features in this energy range. Since we have already ruled
out the fast-cooling regime, there are only two possibilities: the
cooled case where, $\nu_i \sim \nu_c \sim \nu_p \sim 1\ {\rm MeV}$
with a $p\sim 2.4$ and the uncooled case where $\nu_i\sim\nu_p$,
$\nu_c\ga \nu_{\rm max}$ with $p\sim 3.4$. The cooled case can be
ruled out because the energy required in the magnetic field is far too
large. The uncooled case is considered in more detail in the following
paragraphs.

\begin{figure*}
\begin{center}\includegraphics[width=.9\textwidth]{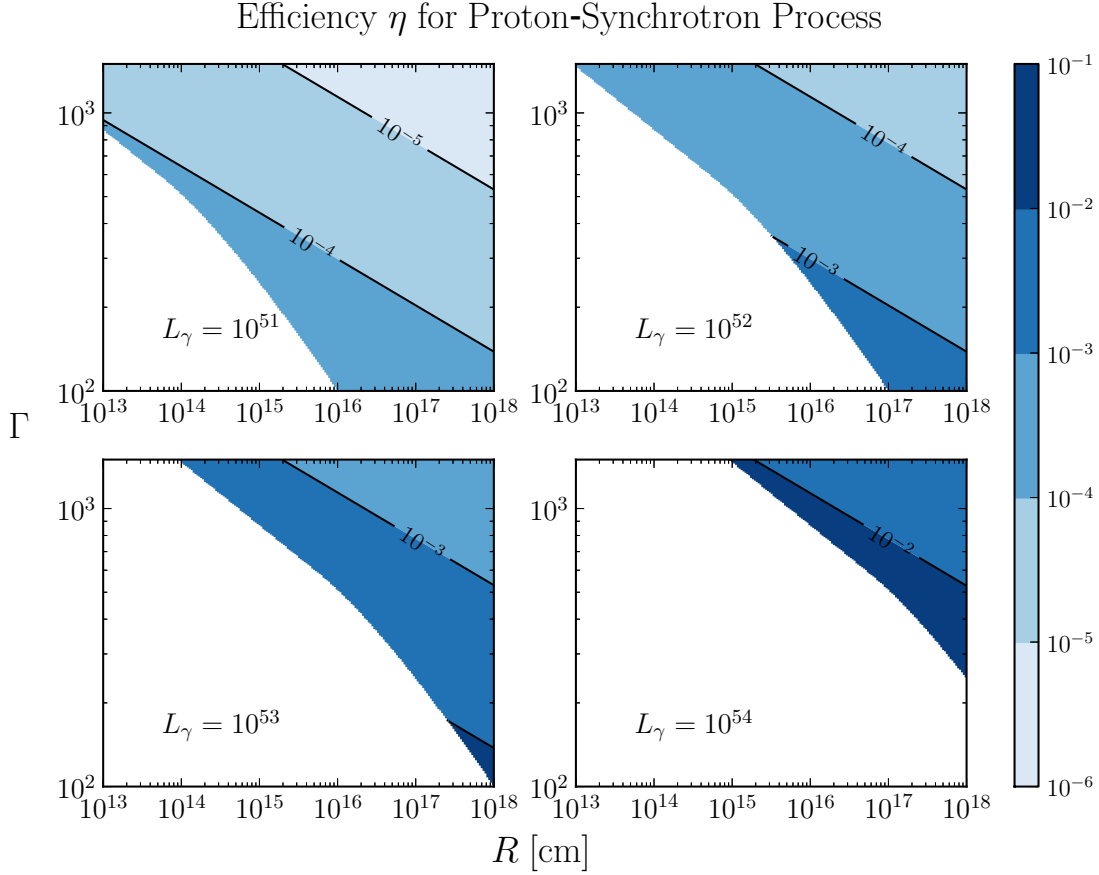}\end{center}
\caption{The maximum $\eta\equiv L_\gamma/(L_p+L_B)$ plotted in the
  $R-\Gamma$ plane for various values of $L_\gamma$ to match a flux of
  $f_\nu=1\ \mu{\rm Jy}$ at 100 MeV. As in the previous figures, we
  fix $z=2$, and $d_L=4.9\times10^{28}$ cm. This $\eta$ corresponds to
  the necessary luminosity in magnetic field and protons with energies
  greater than $\sim4\Gamma$ PeV to match a flux of $f_\nu=1\ \mu{\rm
    Jy}$ at 100 MeV. The luminosity in protons with lower energies is
  likely comparable or much larger. Even when not considering the
  lower energy protons, the efficiency is very small for a vast
  majority of GRB parameter space. When $\eta$ is not shown either the
  place of emission is below the photosphere or it is opaque to
  radiation of 10 GeV due to $\gamma+\gamma$ pair production.}
\label{fig:proton_synch}
\end{figure*}

If we require that $\nu_i\sim\nu_8$ ({\it i.e.} 100 MeV) and
$\nu_c\ga \nu_{\rm max}\sim 10$ GeV, we can then place an upper bound
on the magnetic field by requiring that $\nu_c>\nu_{\rm max}$ and a lower
bound by requiring that protons will be able to be accelerated to high
enough energies to radiate at $\nu_{\rm max}$. As a practical matter,
these bounds do not affect our maximally efficient proton-synchrotron
radiation calculation for the parameter range considered. We then
minimize the total luminosity required in both the magnetic field and
protons radiating at an observed frequency \(\nu_8\) to match a
typical observed flux of a few $\mu$Jy.

The minimum Lorentz factor of the proton that radiates at $\nu_8$ is
\begin{equation}
\gamma_i=4\times10^8B^{-1/2}\Gamma_2^{-1/2}(1+z)^{1/2}\nu_8^{1/2}.
\end{equation}
This Lorentz factor gives a proton luminosity $L_p$ of
\begin{equation}
L_p=2\Gamma^3 \gamma_i R^{-1}m_pc^3N\approx5\times10^{58}f_{\nu,\mu{\rm Jy}}B^{-3/2}\Gamma_2^{3/2}R_{15}^{-1}\nu_8^{1/2}(1+z)^{-1/2}d_{L,28}^2\ {\rm erg/s}.
\end{equation}
Given the magnetic field luminosity, $L_B=6\times
10^{44}B^2R_{15}^2\Gamma_2^2$ erg/s, the total luminosity $L_B+L_p$ will be
minimized with respect to $B$ when $L_p=\frac{4}{3}L_B$, or when
\begin{equation}\label{eq:pro_B}
B\approx10^4\ \Gamma_2^{-1/7}R_{15}^{-6/7}f_{\nu,\mu{\rm Jy}}^{2/7}\nu_8^{1/7}(1+z)^{-1/7}d_{L,28}^{4/7}\ {\rm Gauss}.
\end{equation}
This magnetic field gives a proton luminosity
\begin{equation}
L_p=\frac{4}{3}L_B\approx 5\times10^{52}f_{\nu,\mu{\rm Jy}}^{4/7}\Gamma_2^{12/7}R_{15}^{2/7}\nu_8^{2/7}(1+z)^{-2/7}d_{L,28}^{8/7}\ {\rm erg/s}.
\end{equation}
The efficiency $\eta$ is plotted in figure
  \ref{fig:proton_synch}. Note that $L_B$ is not negligible as in
  figure \ref{fig:efficiency} and \ref{fig:2D}, so in figure
  \ref{fig:proton_synch}, $\eta\equiv L_\gamma/(L_p+L_B)$. The proton
  luminosity is much larger than the $\gamma$-ray luminosity for most
  of the allowed GRB parameter space. Additionally, proton synchrotron
requires an unrealistically large $\gamma_i$. Using eq
(\ref{eq:pro_B}),
\begin{equation}
\gamma_i\approx4\times10^6\ R_{15}^{3/7}\Gamma_2^{-3/7}(1+z)^{4/7}\nu_8^{3/7}f_{\nu,\mu{\rm Jy}}^{-1/7}d_{L,28}^{-2/7}.
\end{equation}
It is unclear what physical process could produce a power law with
such a high minimum Lorentz factor. The minimum Lorentz factor of a
particle accelerated in relativistic shocks is approximately equal
to the Lorentz factor of the shock front with respect to the unshocked
fluid, if every proton crossing the shock front is accelerated. The
Lorentz factor can be proportionally larger if a small fraction of
particles are accelerated and the remaining particles are ``cold''
downstream of the shock front. Considering that the Lorentz factor for
GRB internal shocks is of order a few to perhaps a few tens, the
typical proton Lorentz factor should be $\sim10-10^3$ (the larger
value corresponds to when only 1 in $10^2$ protons are accelerated, as
suggested by simulations, {\it e.g.} 
\citet{2011ApJ...726...75S}). $\gamma_i\sim4\times10^6$
is an unrealistically high injection Lorentz factor for relativistic
shocks. If we set $\gamma_i$ to $10^{3}$, the proton synchrotron
radiation would extend down to $\sim1$ keV and would over produce in
the GBM band. We can only decrease $\gamma_i$ by a factor of 10 before
over producing below the peak of the GRB spectrum. In summary, if
proton synchrotron is to explain the observed LAT emission in GRBs,
all of the protons must be accelerated to extremely high Lorentz
factors very efficiently by some unknown mechanism.

The expected neutrino flux for the proton synchrotron model is
estimated below. The total number of muon neutrinos is
\(\sim2\tau_{p\gamma}N_p\), where \(\tau_{p\gamma}\) is the optical
depth to photo-pion production, given in eq (\ref{eq:optical_depth}),
and $N_p$ is calculated using eq (\ref{eq:pro_flux}) and
(\ref{eq:pro_B}):
\begin{equation}\label{eq:pro_N_nu}
N_\nu=2\times10^{47}L_{\gamma,52}\Gamma_2^{-20/7}R_{15}^{-1/7}f_{\nu,{\rm \mu Jy}}^{5/7}
\nu_{p,6}^{-1}(1+z)^{-13/7}\nu_8^{-1/7}d_{L,28}^{10/7}.
\end{equation}
Because the proton synchrotron radiation requires higher energy protons
and larger magnetic fields compared to the photo-pion process, the
pions produced will suffer larger radiative losses from synchrotron
radiation before they decay. If the magnetic field is given by eq
(\ref{eq:pro_B}) we find that the pions will be cooled significantly
by synchrotron radiation when
\(\Gamma_2^{12/7}R_{15}^{2/7}<4f_{\nu,{\rm\mu Jy}}^{3/7} \nu_8^{5/7}
(1+z)^{2/7} d_{L,28}^{6/7}\). The neutrino flux will peak at an
observed energy of $\sim\frac{1}{4}\Gamma\gamma_im_\pi c^2/(1+z)$ if
the pions are uncooled. If the pions are cooled, the flux will peak at
the energy where pion cooling becomes important,or an energy of
$\sim\frac{1}{4}\Gamma\gamma_{\pi,{\rm cool}}m_\pi c^2/(1+z)$
(\(\gamma_{\pi,cool} \equiv 10^{14}B^{-2}\Gamma_2R_{15}^{-1}\)).
\begin{equation}\label{eq:pro_nu_energy}
E_\nu=\left\{
\begin{array}{ll}\hskip -5pt
1.4\times10^7\ \Gamma_2^{4/7}R_{15}^{3/7}f_{\nu,\mu{\rm Jy}}^{-1/7}
\nu_8^{3/7}(1+z)^{-3/7} d_{L,28}^{-2/7}\ {\rm GeV} & \mbox{if pions
  are not
  cooled}\\ \\
\hskip -5pt 3.5\times10^{6}\Gamma_2^{16/7}R_{15}^{5/7}f_{\nu,\mu{\rm
    Jy}}^{-4/7} \nu_8^{-2/7}(1+z)^{-5/7} d_{L,28}^{-8/7} {\rm GeV} &
\mbox{if pions are cooled}
\end{array}\right.
\end{equation} 
Since the protons are not cooled by the synchrotron loss mechanism, the
observed neutrino flux is calculated using eq (\ref{eq:nu_flux_calc})
and the dynamical time. The neutrino flux, $F_\nu$, peaks at an energy
given by eq (\ref{eq:pro_nu_energy}) and is
\begin{equation}\label{eq:pro_nu_flux}
F_\nu=
\left\{
\begin{array}{ll}\hskip -5pt 1.4\times10^{-3}\ L_{\gamma,52} \Gamma_2^{-2/7} R_{15}^{-5/7} \nu_{p,6}^{-1} \nu_8^{2/7}
f_{\nu,\mu{\rm Jy}}^{4/7}d_{L,28}^{-6/7}(1+z)^{-9/7}\ {\rm
  GeV\ cm^{-2}s^{-1}} & \mbox{if pions are not cooled}\\ \\
\hskip -5pt 3.3\times10^{-4}\ L_{\gamma,52} \Gamma_2^{10/7} R_{15}^{-3/7} \nu_{p,6}^{-1} \nu_8^{-3/7}
f_{\nu,\mu{\rm Jy}}^{1/7}d_{L,28}^{-12/7}(1+z)^{-11/7}\ {\rm
  GeV\ cm^{-2}s^{-1}} & \mbox{if pions are cooled.}
\end{array}\right.
\end{equation}
As in Section \ref{pion_nu}, we estimate the neutrinos detected by
IceCube per second of LAT emission from the proton synchrotron
process:
\begin{equation}\label{eq:pro_nu_count}
\frac{dN_\nu}{dt}=\left\{ \begin{array}{ll}\hskip -5 pt
10^{-3}\ L_{\gamma,52}\Gamma_2^{-4/7}R_{15}^{-13/14}f_{\nu,{\rm \mu
    Jy}}^{9/14}\nu_8^{1/14}\nu_{p,6}^{-1}d_{L,28}^{-5/7}(1+z)^{-15/14}\ {\rm
  counts\ s^{-1}} & \mbox{if the pions are not cooled}
\\ \\ \hskip -5 pt
5.6\times 10^{-4}\ L_{\gamma,52}\Gamma_2^{2/7}R_{15}^{-11/14}f_{\nu,{\rm \mu
    Jy}}^{3/7}\nu_8^{-2/7}\nu_{p,6}^{-1}d_{L,28}^{-8/7}(1+z)^{-17/14}\ {\rm
  counts\ s^{-1}} & \mbox{if the pions are cooled.}
\end{array}\right.
\end{equation}
Therefore, for a bright GRB detected by the Fermi-LAT with
$L_{\gamma,52}\sim10$, $f_\nu,{\rm \mu Jy} \sim 2$, $\Gamma_2 \sim 9$,
$z \sim 2$ and $\nu_{p,6} \sim 1$ and a duration of 10 seconds we find
that the pions will be cooled if $R<2\times10^{14}\ {\rm cm}$. We
expect $\sim 4\times10^{-3} R_{15}^{-13/14}$ neutrinos of energy
$2\times10^{7}R_{15}^{3/7}\ {\rm GeV}$ if $R>2\times10^{14} {\rm
  cm}$ and $\sim 6\times10^{-3} R_{15}^{-11/14}$ neutrinos of energy
$2.7\times10^{7}R_{15}^{5/7}\ {\rm GeV}$ if $R<2\times10^{14}{\rm cm}$.

\section{Summary and Discussion}\label{conclusions}

With a goal of understanding typical observed 100 MeV fluxes in bright
{\it Fermi}-LAT GRBs during the prompt emission, we estimated the
generation of photons by high-energy protons traveling through a shell
of photons whose energy distribution is given by the Band function. We
calculated the minimum energy in protons required to reproduce {\it
  Fermi}-LAT observations for the following hadronic processes:
photo-pion, Bethe-Heitler pair production, and proton synchrotron.

Unlike previous works, we specifically focused on the energy required
for hadronic models to produce the \(>\)100 MeV photons seen in {\it
  Fermi} GRBs and how this requirement depends on GRB parameters. To
provide additional physical insight into the energy requirements, we
have provided both analytical estimates and more detailed numerical
calculations.

We find that photo-pion $\Delta$-resonance is much more efficient than
Bethe-Heitler pair production at producing high-energy electrons---so
much so that Bethe-Heitler pair production can be ruled out as a
mechanism for producing $\ga 100$ MeV photons observed by {\it
  Fermi}-LAT. The photo-pion process is capable of producing high
energy electrons, but to match the {\it Fermi}-LAT flux at 100 MeV,
the photo-pion process requires an energy in protons that is $\ga
10^4$ times greater than isotropic energy in the $\gamma$-ray
photons. Since the Bethe-Heitler photo-pairs are produced more
efficiently at low energies, Bethe-Heitler production will be the
dominant process at low energies. These low-energy Bethe-Heitler
electrons will have the same spectral index as the high-energy
photo-pion electrons (assuming there isn't a cooling break). Therefore
it is possible that both processes could add up to produce a single
power-law deviation from the Band function that extends from low to
high energies. This type of spectral feature has been observed in
several {\it Fermi} GRBs.

According to our calculations, proton synchrotron is capable of
producing the {\it Fermi}-LAT GRB emission more efficiently than the
other hadronic processes. The proton synchrotron could possible
achieve efficiencies on the order of 1-10\% for the brightest GRBs, if
we assume that the minimum Lorentz factor for proton accelerated in
shocks is extremely large $\sim2\times 10^6$. The minimum proton
Lorentz factor of order $10^6$ is much larger than what is expected
based on our current understanding of relativistic collisionless
shocks. Regardless of the mechanism accelerating the protons, these
high energy protons with LF greater than $10^6$ likely carry only a
small fraction of the total energy carried by the protons.

We also calculated the expected neutrino flux if the LAT emission is
from the photo-pion process or proton synchrotron radiation. In the
photo-pion process, for a bright LAT GRB with a Lorentz Factor of 900
and a duration of 10 seconds, we expect $\sim .1$ neutrinos detected
by IceCube at an energy of $\sim10^6\ {\rm GeV}$. Therefore, it may be
possible to rule out photo-pion emission when the emission
from multiple bursts is considered. For proton synchrotron radiation,
the neutrino flux also depends on the emission radius, $R$. For a
bright LAT GRB with a Lorentz factor of 900, an emission radius of
$10^{15}$ cm and duration of 10 seconds, we expect a $\sim
4\times10^{-3}$ neutrinos detected by IceCube at an energy
$\sim2\times10^{7}\ {\rm GeV}$.

In summary, all the hadronic processes considered in this paper
require significantly more energy in protons than the observed energy
in gamma-rays to reproduce the high-energy flux observed in {\it
  Fermi}-LAT GRBs.

\section{Acknowledgments}
This work has been funded in part by NSF grant ast-0909110, and a
Fermi-GI grant (NNX11AP97G). Patrick would like to thank
Rodolfo Barniol Duran and Rodolfo Santana for their helpful discussions and
his wife Diana for her support and help preparing the paper.

\bibliographystyle{mn2e}
\bibliography{paper.bib}

\end{document}